\documentclass[showpacs,floatfix,twocolumn,secnumarabic,amssymb, nobibnotes, aps, prd, nofootinbib]{revtex4-1}

\usepackage{amssymb,amsmath,amsfonts}
\usepackage{array}

\usepackage[inline]{enumitem}
\usepackage{nicefrac}
\usepackage{braket}
\usepackage{fullpage}
\usepackage{fancyhdr}
\usepackage{graphicx}
\usepackage{amsthm,multibox}
\usepackage{color}
\usepackage{hhline,xspace,rotate}
\usepackage{epsfig}
\usepackage{epstopdf}
\usepackage{pst-all}
\usepackage{pst-poly}
\usepackage{pst-coil}
\usepackage{multido}
\usepackage{url}
\usepackage[french]{varioref}
\usepackage{colortbl}
\usepackage{verbatim}
\usepackage{placeins}
\usepackage{appendix}
\usepackage{dsfont}
\usepackage{latexsym}

\usepackage{setspace}
\usepackage[breaklinks=true,debug=true]{hyperref}
\usepackage[utf8]{inputenc}
\usepackage[T1]{fontenc}
\usepackage[frenchb, english]{babel}

\usepackage[para,online,flushleft]{threeparttable}

\begin{document}

\title{Radial Overlap Correction to Superallowed $0^+ \to 0^+$ $\beta $-decay revisited}%

%\title{Radial Overlap Correction from the Shell Model with Woods-Saxon Potential}

\author{L. Xayavong}%
%\email{xayavong@cenbg.in2p3.fr}
\author{N. A. Smirnova}%
%\email{smirnova@cenbg.in2p3.fr}
\affiliation{CENBG (CNRS/IN2P3 -- Université de Bordeaux),
33175 Gradignan cedex, France}
\date{\today}%

\begin{abstract}
Within the nuclear shell model, we investigate the correction $\delta_{RO}$ to the Fermi matrix element 
due to a mismatch between proton and neutron single-particle radial wave functions.  
Eight superallowed $0^+ \to 0^+$ $\beta$ decays in the $sd$-shell, comprised of 
$^{22}$Mg, $^{26m}$Al, $^{26}$Si, $^{30}$S, $^{34}$Cl, $^{34}$Ar,
$^{38m}$K and $^{38}$Ca are re-examined. 
The radial wave functions are obtained from a spherical Woods-Saxon potential 
whose parametrizations are optimized in a consistent adjustment of 
the depth and the length parameter to relevant experimental observables, 
such as nucleon separation energies and charge radii, respectively. 
The chosen fit strategy eliminates the strong dependence of the radial mismatch correction to a specific parametrization, 
except for calculations with an additional surface-peaked term. As an improvement, 
our model proposes a new way to calculate the charge radii, based on a parentage expansion which accounts 
for correlations beyond the extreme independent-particle model. Apart from the calculations with a surface-peak term 
and the cases where we used a different model space, the new sets of $\delta_{RO}$ are in general agreement 
with the earlier result of Towner and Hardy~\cite{ToHa2002}. 
Small differences of the corrected average $\overline{\mathcal{F}t}$ value are observed.

\end{abstract}

\pacs{21.60.Cs, 23.40.Bw, 23.40Hc, 27.30.+t}

\maketitle
%\tableofcontents

\section{Introduction}\label{Introduction}
The superallowed nuclear $\beta$ decay between $0^+$, $T=1$ isobaric analog states  
has long been known as a sensitive tool to probe the fundamental symmetries underlying
the Standard Model of electroweak interaction. According to the Conserved Vector Current~(CVC) hypothesis the 
corrected $\mathcal{F}t$ value or equivalently the vector coupling constant $G_V$ must be nucleus-independent. 
If CVC holds, those constants can be used to extracted $|V_{ud}|$,
the absolute value of the up-down element of the Cabibbo-Kobayashi-Maskawa~(CKM) quark-mixing matrix. 
The last one, combined with the complimentary experimental data on $|V_{us}|$ and $|V_{ub}|$, the other top-row elements, 
provides the most accurate test of the unitarity of the CKM matrix 
(see Ref.~\cite{HaTo2015} for details and the present status). 

At present, 14 transitions ranging from $^{10}\mbox{C}$ to $^{74}\mbox{Rb}$ are known experimentally 
with a precision of 0.1\% or better, therefore, we must consider all kinds of side effects of this order of magnitude 
and estimate the necessary corrections before deducing the $\mathcal{F}t$ value.  
All previous investigations (Ref.~\cite{HaTo2015} and references therein) indicate that the current uncertainty on $|V_{ud}|$ 
is dominated by a set of theoretical corrections aimed to account for the radiative effects and 
the isospin-symmetry breaking in nuclear states. 
The latter is strongly structure-dependent and has the greatest effect on reducing the scatter in the $\mathcal{F}t$ values.

Since superallowed $0^+ \to 0^+$ $\beta $ decay is governed uniquely by the vector part of the electroweak current,
the corrected $\mathcal{F}t$ value can be deduced from the expression~\cite{ToHa2008}, 
\begin{eqnarray}\label{1}
\mathcal{F}t  & = ft(1+\delta_{R}')(1+\delta_{NS}-\delta_C) \\ \nonumber
  & = \displaystyle \frac{K}{2G_V^2(1+\Delta_R^V)}=const,
\end{eqnarray}
where $K$ is a combination of fundamental constants 
$K=2\pi^3\hbar \ln2 (\hbar c)^6/(m_e c^2)^5=(8120.2716\pm 0.012)\times10^{-10}~\mbox{GeV}^{-4}\mbox{s}$, 
$ft$ is the product of the statistical rate function ($f$)~\cite{f} and the partial half-life ($t$). 
The radiative corrections are separated into three parts~\cite{HaTo2015} : 
$\Delta_R^V= 2.361(38)\%$ is nucleus independent, 
$\delta_R'$ depend on the atomic number of daughter nucleus and 
$\delta_{NS}$ is nuclear-structure dependent. 
The correction due to the isospin-symmetry breaking, $\delta_C$, is defined as the deviation of 
the realistic Fermi matrix element squared from its isospin-symmetry value, 
\begin{equation}\label{2}
|M_F|^2 = |M_F^0|^2(1-\delta_C),
\end{equation}
with $|M_F^0|=\sqrt{T(T+1)-T_{zi}T_{zf}}=\sqrt{2}$ for the $T=1$ case. 
%The relationship between the Fermi and the vector coupling constant is : $G_V=G_F|V_{ud}|$, 
%where $G_F$ is the Fermi coupling constant, equal to $(\hbar c)^3 \times 1.166 378 7(6) \times 10^{-5}\mbox{GeV}^{-2}$.

Over the past 40 years, numerous theoretical work has been devoted to $\delta_C$  
within various approaches. For example, Towner and Hardy~\cite{ToHa2002,ToHa2008,HaTo2009} use 
a nuclear shell model with radial wave functions derived from a Woods-Saxon (WS) potential. 
The calculations based on the density functional theory with a spin and isospin projected technique 
have been realized by Satula et al~\cite{Sat2011}. 
Another approach in which RPA correlations have been added to a relativistic Hartree or Hartree-Fock (HF) calculation 
was used by Liang et al~\cite{Li2009}. 
In addition, Auerbach~\cite{Au2009} uses a model where the main isospin-symmetry-breaking effects are attributed 
to the isovector monopole resonance. 
All those calculations lead to a significant spread in the obtained values of $\delta_C$, 
raising thus the question of credibility of the results.

The values for $\delta_C$ tabulated by Towner and Hardy in Ref.~\cite{HaTo2015} excellently support 
both the CVC hypothesis over the full range of $Z$ values and  the top-row unitarity of the CKM matrix. 
However, this agreement is not sufficient to reject the other calculations, 
since these aspects of the Standard Model could not yet be confirmed experimentally. 
The confirmation of CVC does not constrain the absolute  $\mathcal{F}t$ value. 
The dissensus between model predictions and the importance of the issue motivated us to re-examine this correction 
in a consistent approach.

Within the nuclear shell model~\cite{OrBr1985,ToHa2008}, the correction $\delta_C$ can be represented with good accuracy 
as a sum of two terms~\footnote{The two terms are 
referred to as $\delta _{C1}$ and $\delta_{C2}$ in the work of Towner and Hardy.}: 
\begin{equation}\label{3}
\delta_C \approx  \delta_{IM} + \delta_{RO} \,.
\end{equation}
The first term on the right-hand side ($\delta_{IM}$) accounts for the isospin mixing among $0^+$ state wave functions in the
parent and daughter nuclei. 
This part is calculated from configuration mixing within a shell-model calculation using charge-dependent interactions. 
The details of the calculations for $\delta_{IM}$ are described in
Refs.~\cite{ToHa2008, OrBr1985,OrBr1995,YiHua}. 
The last term ($\delta_{RO}$) in Eq.~\eqref{3} corrects for the deviation from unity of the overlap integral 
between the radial part of proton and neutron single-particle wave functions. 
The protons in the parent nucleus are typically less bound than the neutrons in the daughter nucleus because 
of the Coulomb repulsion. 

Currently, two types of a mean-field potential are considered for evaluating the correction $\delta_{RO}$. 
The first one is the phenomenological WS potential including a central, a spin-orbit and an electrostatic repulsion terms. 
A series of calculations using this potential has been carried out by Towner and Hardy~\cite{ToHa2002,ToHa2008}. 
These authors adjusted case-by-case the depth of the volume term 
or added an additional surface-peak term to reproduce experimental proton and neutron separation energies. 
In addition, they adjusted the length parameter of the central term to fix the charge radius of the 
parent nuclei. 
The second type of a mean-field potential is that obtained from self-consistent HF calculations using a zero-range Skyrme force, 
as was first proposed by Ormand and Brown in 1985~\cite{OrBr1985}. 
A more completed compilation for $\delta_{RO}$ from these authors were published later~\cite{OrBr1989,OrBr1995}. 

The values for $\delta_{RO}$ obtained with both types of a mean-field potential are equivalently
in good agreement with the CVC hypothesis, 
however, those resulted from Skyrme-HF calculations  
are consistently smaller than the others. This discrepancy has been commonly understood as 
insufficiency of the Slater approximation for treating the Coulomb exchange term. 
Towner and Hardy highlighted that the asymptotic limit 
of the Coulomb potential in the Slater approximation is overestimated by one unit of $Z$. 
To retain this property, they proposed a modified HF protocol~\cite{HaTo2009}, namely they performed a single calculation for
the nucleus with $(A-1)$ nucleons and $(Z-1)$ protons and then using 
the proton and the neutron eigenfunctions from the same calculation to
compute the radial overlap integrals. Their result leads to a significant increase of $\delta_{RO}$ 
and provides a better agreement with the values obtained with WS eigenfunctions. 
However, this method is rooted to the Koopman's theorem which 
is not fully respected by such HF calculations, in particular, with a density-dependent effective interaction. 

In the present paper, we propose a comprehensive and detailed study of the radial-overlap correction
to superallowed $0^+ \to 0^+$ $\beta $-decay matrix elements 
using the nuclear shell model with WS single-particle wave functions.
A special emphasis is given on the choice of the WS potential parametrization and optimization procedure. 
We limit ourselves to the $sd$-shell nuclei, for which very precise shell-model wave functions are available.
Once the method is established, we plan to extend this study to heavier emitters, 
using large-scale shell-model diagonalization and modern effective interactions.

The article is organized as follows.  
The general formalism is given in section~\ref{formalism}. 
Section~\ref{ws} is devoted to the selection of a WS potential parametrization with brief discussion of 
physics aspects behind the construction.
In section~\ref{overlap}, we carry out a simplified calculation of $\delta_{RO}$ 
without taking into account the sum over intermediate states. 
The sensitivity to the choice of the parametrization and adjustment procedure is investigated. 
In section~\ref{interm}, we present our final results on the correction,
obtained from a full parentage expansion for both $\delta_{RO}$ 
correction and charge radii of the parent nuclei.
%so that the whole spectrum of intermediate states could be allowed to be operative. 
The charge radii are computed using two different methods with respect to the treatment of closed-shell orbits. 
In section~\ref{Const}, we use the results for $\delta_{RO}$ to get the weighted averages 
of the $\mathcal{F}t$ values for six $sd$-shell emitters for which 
the measured $ft$ values have attained the level of precision currently required for the tests of the Standard Model. 
In section~\ref{CVC}, the sets of $\delta_{RO}$ are tested against the experimental data, 
under the assumption that the CVC hypothesis is valid. 
Comparison with the previously published values are made. 
The summary and concluding remarks are given in section~\ref{Summary}.

\section{General formalism}\label{formalism}

Even though the basic formulas we are going to use are already described by several 
authors~\cite{OrBr1985,ToHa2008,MiSch2008,MiSch2009}, it is worth summarizing 
them here.
%so that it is clear exactly how we use them and to set the notation for the following sections.  

We start from the exact formalism developed in Refs.~\cite{MiSch2008,MiSch2009} which 
uses the basis states given by a realistic mean-field potential with presence of   
isospin non-invariant terms. A mismatch between neutron and proton 
radial wave functions leads to the nodal-mixing relation: 
\begin{equation}\label{node}
a_{\beta_p }^{\dagger} = \sum_\alpha a_{\alpha_n }^{\dagger} \braket{\alpha_n|\beta_p}, 
\end{equation}
where we use the subscript $p$ to indicate proton states and $n$ for neutron states, and 
the sum is over the radial quantum number only. 
The operator $a_{\alpha}$ destroys a nucleon in quantum state $\alpha $ whereas the operator 
$a_{\alpha}^{\dagger}$ creates a nucleon in that state,
with $\alpha $ standing for the whole set of spherical quantum numbers: 
$\alpha =(n_{\alpha }, l_{\alpha }, j_{\alpha }, m_{\alpha })$. 
The overlap matrix in Eq.~\eqref{node} is given by: 
\begin{eqnarray}\label{5}
\braket{\alpha_n|\beta_p} & = & \displaystyle \delta'_{\alpha \beta} \times \Omega_{\alpha\beta} , 
\end{eqnarray}
where 
\begin{equation}
\Omega_{\alpha\beta} =  \int_0^\infty R_{\alpha_n}(r) R_{\beta_p}(r) r^2 dr
\end{equation}
and the Kronecker delta, $\delta'_{\alpha \beta}$, represents the orthogonality of the angular part of 
single-particle wave functions due to the spherical symmetry, that is 
$\delta'_{\alpha \beta} = \delta_{l_{\alpha} l_{\beta}} \times \delta_{j_{\alpha} j_{\beta}} \times  \delta_{m_{\alpha} m_{\beta}}$. 
The symbol $R_{\alpha_n}(r)$ and $R_{\beta_p}(r)$ denote single-particle radial wave functions of a neutron and a proton respectively, 

Within Eq.~\eqref{node}, the nuclear matrix element for 
the superallowed $\beta^+$-decay can be expressed as 
\begin{equation}\label{MF}
\begin{array}{ll}
M_F & = M_F^{TH} + \delta M_F , 
\end{array}
\end{equation}
where $M_F^{TH}$ has been referred to as the Towner-Hardy term and $\delta M_F$ is 
an interference term as suggested by Miller and Schwenk~\cite{MiSch2009}. 
These terms are given by 
\begin{equation}\label{MTH}
\begin{array}{ll}
 M_F^{TH}  & = \displaystyle \sum_{\alpha }\bra{f} a_{\alpha_n }^{\dagger} a_{\alpha_p }\ket{ i} \bra{\alpha_n} t_+\ket{\alpha_p}, 
\end{array}
\end{equation}
and 
\begin{equation}\label{MMS}
\begin{array}{ll}
\delta M_F & =  \displaystyle \sum_{\alpha,\beta }^{\alpha\ne\beta}\bra{f} a_{\alpha_n }^{\dagger} a_{\beta_p }\ket{ i} \bra{\alpha_n} t_+\ket{\beta_p}, 
\end{array}
\end{equation}
where $t_+$ is the exact isospin raising operator which satisfies the isospin
commutation relations (see discussion in Ref.~\cite{MiSch2009}), and 
$\ket{i}$ and $\ket{f}$ denote the initial and the final nuclear states, respectively. 
Here, we use the proton-neutron formalism, therefore in coordinate representation the operator $t_+$ 
is equivalent to the identity operator, thus the single-particle matrix elements are reduced to 
\begin{equation}
\bra{\alpha_n} t_+\ket{\beta_p} = \braket{\alpha_n|\beta_p}
\end{equation}

Here we notice that it is not possible to include the interference term within the shell model 
because of the nodal mixing which requires a large model space (at least 2 oscillator shells). 
For this reason, we will stay within the same approximation as Towner and Hardy~\cite{ToHa2002,ToHa2008} 
which considers only the diagonal term. 

It has been pointed out by Miller and Schwenk~\cite{MiSch2009} on the basis of a schematic model
that both terms in Eq.~\eqref{MF} are of the same order of magnitude. 
No realistic calculation have been performed yet.
%however, these authors do not produce any exact calculations,  
%instead they proceed to make a model assumptions of their own, from which they conclude 
%without numerical results. 
Although the model of Towner and Hardy lacks the interference term, 
the calculated $\delta_C$ values eliminate much of the important scatter 
present in the uncorrected $ft$ values. 
This may be considered as an evidence that the terms omitted by
Towner and Hardy must be either nucleus independent or must have a negligible effect.

Let us define the correction to the one-body transition density due to the isospin-symmetry breaking as 
\begin{equation}
\Delta_\alpha = \bra{f} a_{\alpha_n }^{\dagger} a_{\alpha_p }\ket{ i}^T-\bra{f} a_{\alpha_n }^{\dagger} a_{\alpha_p }\ket{ i}, 
\end{equation} 
where the superscript $T$ is used to denote the one-body transition densities evaluated at the isospin-symmetry limit. 

One can express the matrix element $M_F^{TH}$ 
in terms of the correction $\Delta_\alpha$ as 
\begin{eqnarray}\label{MF1}
\displaystyle M_F^{TH}  & =& \sum_{\alpha } \big( \bra{f} a_{\alpha_n }^{\dagger} a_{\alpha_p }\ket{ i}^T - \Delta_\alpha \big)  \Omega_\alpha \\\nonumber
        & =& \sum_{\alpha } \big( \bra{f} a_{\alpha_n }^{\dagger} a_{\alpha_p }\ket{ i}^T - \Delta_\alpha \big)  \big[1-(1-\Omega_\alpha)\big] \\\nonumber
        & =& M_F^0 \big[ 1-\frac{1}{M_F^0}\sum_\alpha\Delta_\alpha + \frac{1}{M_F^0}\sum_\alpha \Delta_\alpha(1-\Omega_\alpha)  \\\nonumber
        &-& \frac{1}{M_F^0}\sum_\alpha \bra{f} a_{\alpha_n }^{\dagger} a_{\alpha_p }\ket{ i}^T (1-\Omega_\alpha)   \big], 
\end{eqnarray}
where $\Omega_\alpha=\Omega_{\alpha\alpha}$. 

Thus, the matrix element squared is 
\begin{eqnarray}\label{MF2}
\displaystyle|M_F^{TH}|^2  & =& |M_F^0|^2 \Big[ 1 -\frac{2}{M_F^0}\sum_\alpha\Delta_\alpha  \\
        &-& \frac{2}{M_F^0}\sum_\alpha \bra{f} a_{\alpha_n }^{\dagger} a_{\alpha_p }\ket{ i}^T (1-\Omega_\alpha) + \mathcal{O}(\zeta^2)  \Big], \nonumber
\end{eqnarray}
where $\zeta$ denotes $(1-\Omega_\alpha)$ or $\Delta_\alpha$. 

Since the isospin-symmetry-breaking effect is small, it is convenient to neglect higher-order terms of Eq.~\eqref{MF2}.  
In this way one obtains a suitable expression for $\delta_C$ as given in Eq.~\eqref{3}, with the radial overlap part : 
\begin{equation}\label{RO}
\displaystyle \delta_{RO} = \frac{2}{M_F^0}\sum_\alpha \bra{f} a_{\alpha_n }^{\dagger} a_{\alpha_p }\ket{ i}^T (1-\Omega_\alpha), 
\end{equation}
and the isospin-mixing part : 
\begin{equation}\label{IM}
\displaystyle \delta_{IM} = \frac{2}{M_F^0}\sum_\alpha\Delta_\alpha. 
\end{equation}

The initial and final state wave functions will be determined by diagonalization of a well-established 
shell-model effective Hamiltonian in a spherical (harmonic-oscillator) many-body basis. 
With addition of isospin non-conserving terms, one can obtain isospin-mixed wave functions that can be used 
to compute $\delta_{IM}$. 
The contribution  $\delta_{RO}$  beyond the model space is accounted for
by the overlap integrals, $\Omega_\alpha$, which slightly deviate from unity when 
evaluated with realistic radial wave functions.

\section{Woods-Saxon potential}\label{ws}

%Among existing in literature phenomenological
%potentials, the WS potential is the
%best to describe the individual motion of nucleon inside the nucleus. It
%provides a model for the properties of bound states as
%well as the continuum single-particle wave functions,
%which can be served as a realistic single-particle basis
%for shell-model calculations. 

The standard form of a WS potential is based upon the sum of a spin-independent
central term, a spin-orbit term, an isospin-dependent term, and a term that accounts for the Coulomb repulsion: 
\begin{eqnarray}\label{WS}
\displaystyle V(r)&=& + V_{0}f(r,R_0,a_0)  \nonumber \\
\displaystyle && + V_{s}\left(\frac{r_s}{\hbar}\right)^2 \frac{1}{r}\frac{d}{dr}[f(r,R_s,a_s)]\braket{\boldsymbol{l}\cdot \boldsymbol{\sigma}} \\ \nonumber
\displaystyle && + V_{iso}(r)  \\ \nonumber
\displaystyle && + V_{c}(r) \,,
\end{eqnarray}
where
\begin{equation}\label{fermi}
\displaystyle f(r,R_i,a_i)=\frac{1}{1+\exp{(\frac{r-R_i}{a_i})}} \, ,
\end{equation}
with $i$ denoting either $0$ for the central term (the first term on the the right-hand side of Eq.~\eqref{WS}) or 
$s$ for the spin-orbit term (the second term on the right-hand side of Eq.~\eqref{WS}). 
The radius, $R_i$, and the diffuseness of the surface, $a_i$, are fixed parameters. % of the same units of length as $r$. 

The independent-particle model utilizing the potential given in Eq.~\eqref{WS} cannot be 
solved analytically, therefore it is not possible to separate out the spurious center-of-mass contribution.  
To the authors' knowledge, the most practical way for eliminating such a contribution 
is to adopt artificially the nucleon-core/target concept from the optical model~\cite{Hod1971}. 
Based on that two-body picture, the one-body Schrödinger equation could be
solved in relative coordinates by simply replacing the nucleon mass, $m$ with the reduced mass:
\begin{equation}\label{mass}
\displaystyle \mu = m \times \frac{A-1}{A} \,, 
\end{equation}
where $A$ is the mass number of the composite nucleus. 

Note also that all terms in Eq.~\eqref{WS} are local, this corresponds to 
the Hartree approximation of the self-consistent mean-field theory. 
We therefore encounter the self-interaction effects due to the lack of exchange terms. 
In general, this problem is remedied with the assumption that 
the mean-field potential being generated by only the $(A-1)$ nucleons of the core nucleus,
neglecting thus the contribution of the nucleon of interest.  
Within this restriction, any mass dependence should be expressed in terms 
of $(A-1)$ instead of $A$. Thus, the radii read  
\begin{equation}
R_i=r_i\times (A-1)^{1/3}\,. 
\end{equation}

%Let us now highlight the physics foundations behind the construction of the terms given in Eq.~\eqref{WS}. 
The Woods-Saxon form~\eqref{fermi} decreases exponentially with increasing radius. This is
in strong agreement with the fact that the density of nucleons in the nucleus is fairly constant in the interior
and drops smoothly to zero beyond the nuclear radius~\cite{DG}, and 
further it is efficient in satisfying the saturation features of nuclear forces. 

The spin-orbit term is taken to have the Thomas form~\cite{Thomas}, similar to that used for 
electrons in atoms which was derived from the Dirac theory. 
Although it takes a similar form, the nuclear spin-orbit coupling is not a relativistic correction, 
but a first order term in the bare nucleon-nucleon interaction. It is well known that such a force 
is responsible for the shell structure of nuclei and of the opposite sign to that in atoms. 
In general, the spin-orbit radius ($r_s$) is smaller than the radius ($r_0$) of the volume term, because of 
the very short range of the two-body spin-orbit interaction~\cite{BohrMott}. 

%The nuclear spin-orbit force manifests itself mainly in the surface region of nuclei. 
%In the nuclear interior, where the nucleon density is approximately constant, 
%there are an equal number of nucleons on either side of 
%the orbital, the spin-orbit interaction consequently averages out. 
%That is why the form factor the spin-orbit potential is given by the first derivative of 
%the WS form as given in Eq.~\eqref{WS}. In general, the spin-orbit radius ($r_s$) is smaller than the central radius ($r_0$), because of 
%the very short range of the two-body spin-orbit interaction \cite{BohrMott}. 
%This is why $r_s$ is typically chosen to be smaller than $r_0$. 

There is numerous experimental evidence~\cite{LanePRL1962,Lane1962,Hod1971} from both positive and negative energies for the shift
between the nuclear part of neutron and proton potentials. To account for this effect, we have 
to add to the nuclear single-particle potential an isospin-dependent term. 
A number of authors~\cite{Satchler,GS} have proposed a simple form which tends to favor a
balanced configuration of neutrons and protons, namely
\begin{equation}\label{symm}
V_{iso}(r) = \displaystyle V_1 \frac{t_z\cdot T_z'}{A-1} f(r,R_0,a_0), 
\end{equation}
where $t_z$ is the isospin projection of the nucleon, with $t_z=1/2$ for neutron and $-1/2$ for proton, and 
$T_z'$ is the isospin projection of the core/target nucleus. 
The form factor in Eq.~\eqref{symm} is assumed to have the same form as the volume term. 

It was pointed out later by Lane~\cite{LanePRL1962,Lane1962} that the symmetry term~\eqref{symm} is an averaged version of
a more fundamental formula which contains a dependence on
the scalar product of the isospin operators of a nucleon ($\boldsymbol{t}$) and a core nucleus ($\boldsymbol{T}'$) 
(see also discussion in Ref.~\cite{SWV}): 
\begin{equation}\label{iso}
V_{iso}(r) = \displaystyle V_1 \frac{ \braket{\boldsymbol{t}\cdot \boldsymbol{T}'}}{A-1} f(r,R_0,a_0). 
\end{equation}

In principle, we could include all extended symmetry preserving terms which involve the nucleon operators 
$\boldsymbol{p}$, $\boldsymbol{r}$, $\boldsymbol{\sigma}$, $\boldsymbol{t}$ and the core
spin and isospin operators $\boldsymbol{T}'$ and $\boldsymbol{J}'$. 
However, most of these terms were found to be small~\cite{Hod1971}, only the isospin dependence of the spin-orbit strength,  
i.e. $\braket{\boldsymbol{T}'\cdot \boldsymbol{t}}\braket{\boldsymbol{l}\cdot \boldsymbol{\sigma}}$, must be included 
for study of neutron-rich nuclei. This term can be parametrized as~\cite{ISOor,ISOor1} 
\begin{equation}\label{iso-ls}
\begin{array}{ll}
V_{iso}^s(r)  & = \displaystyle V_1^s \frac{ \braket{\boldsymbol{t}\cdot \boldsymbol{T}'}}{A-1} 
\left(\frac{r_s}{\hbar}\right)^2 \frac{1}{r} \frac{d}{dr}[f(r,R_s,a_s)]\braket{\boldsymbol{l}\cdot \boldsymbol{\sigma}}.
\end{array}
\end{equation} 

Normally, the strength of the spin-orbit term is related to that of the volume term by, 
\begin{equation}
V_s = -\lambda\times V_0, 
\end{equation}
and for the isospin-dependent part, 
\begin{equation}\label{1s}
V_1^s = -\lambda_1\times V_1. 
\end{equation}

The depth of each term in Eq.~\eqref{WS} depends in general on momentum, 
reflecting the nonlocal nature of the nuclear potential~\cite{LanePRL1962,Lane1962,Hod1971}. 
However, the higher order terms are
usually neglected and only the zero order term (a constant) is taken into account. 

At last, the repulsive long-range Coulomb potential is determined from the assumption of a uniformly charged sphere of a radius $R_c$. 
For numerous applications, this is a good approximation because the influence of the surface diffuseness 
of the charge distribution on the strength of the Coulomb potential is not strong. 
The well-known analytical expression reads: 
\begin{equation}\label{12}
V_c(r)=(Z-1)e^2\times\left \{
\begin{array}{ll}
\displaystyle\frac{1}{r}, & \mbox{if $r>R_c$}; \\
\displaystyle\frac{1}{R_c}\left(\frac{3}{2}-\frac{r^2}{2R_c^2}\right), & \mbox{otherwise} \,. 
\end{array}
\right.
\end{equation}
In general, the parameter $R_c$ is defined in the same way as the central and spin-orbit radii: $R_c=r_c\times (A-1)^{1/3}$. 
However, since the Coulomb term is of major interest for our purpose, 
we will extract the parameter $R_c$ from experimental data on charge radii, $\braket{r^2}_{ch}$ via the following formula~\cite{Elton}:
\begin{equation}\label{Rc}
R_c^2 = \frac{5}{3} \bigg[ \braket{r^2}_{ch} - \frac{3}{2}\left(a_p^2-b^2/A\right) \bigg] \, .
\end{equation}
In this equation, the last two terms correct for the internal structure of the proton and for the center-of-mass motion, 
with $a_p=0.694$ fm~\cite{ToHa2002} being the parameter of the Gaussian function describing the charge distribution of the proton and 
$b$ being the harmonic-oscillator length parameter. 

There exist in literature a number of the WS potential parametrizations
(Refs.~\cite{BohrMott,LoDu1998,SWV, WS1, WS2, WS3} and references therein), 
constructed with different objectives and relevant for different nuclear mass regions. 
In this work, we select two parametrizations that seem to us to
be appropriate for our purposes. 
One of them is that of Bohr and Mottelson~\cite{BohrMott}, modified as proposed in Ref.~\cite{Xthesis} 
and denoted as BM$_m$, while the other is that of Schwierz, Wiedenhöver and Volya (SWV), published in Ref.~\cite{SWV}. 
They differ mainly by the isovector term which represents the second source of the isospin-symmetry violation in a WS potential. 
The parametrization BM$_m$ includes the symmetry term of the form Eq.~\eqref{symm}, 
whereas the other employs the isospin coupling as given in Eq.~\eqref{iso}. 
For heavy nuclei with large neutron excess, the difference is small. %~\cite{SWV}. 
However, it leads to significantly different predictions in lighter nuclei around $N=Z$ line, which is the region of primary interest of our study. 
Note also that in the SWV parametrization, the radii of the central and the spin-orbit terms are calculated with respect to the composite nucleus, i.e.
$R_0=r_0 \times A^{1/3}$ and $R_s=r_s \times A^{1/3}$, respectively. 

The numerical values of both parametrizations are summarized in Table~\ref{para}. 

\begin{table}[h!]
\centering
\caption{Numerical values of the selected parametrizations.}
\setlength{\extrarowheight}{0.08cm}
\begin{tabular}{| p{1.3cm} | p{1.5cm} |p{2.8cm}|p{1.3cm}|}
\hline
           & BM$_m$         & SWV  & Unit          \\
\hline%\hline
  $r_0$    & 1.26                   & 1.26    &         fm                   \\
  $r_s$    & 1.16                   & 1.16       &          fm               \\
  $a_0=a_s$& 0.662                  & 0.662   &           fm                 \\
  $V_0$    & $-52.833$                 & $-52.06$     &               MeV            \\
  $V_1$    & $-146.368$                &  $-133.065$  &                MeV               \\
  $\lambda$ & 0.22               &  $0.198 A^2/(A-1)^2$    &       -                   \\
  $\lambda_1$ & 0.22               &  0.          &          -          \\
% Coulomb    & charge uniform sphere         & charge uniform sphere                            \\
\hline
\end{tabular}
\label{para}
\end{table}

\section{Radial Overlap Correction in the closure approximation}\label{overlap}  

To date, all calculations of $\delta_{RO}$ include a refinement that accommodates the whole spectrum 
of the intermediate $(A-1)$ nucleus. 
However, it is instructive to first consider the simplest approach which assumes  
that the ground state of the $(A-1)$ nucleus is a unique parent. 
A calculation of this type would be too crude to produce quantitative result for $\delta_{RO}$. 
Rather, our purpose here is to study the parametrization dependence and to see how the correction will be changed 
when the calculation is carried out with the full parentage expansion, as will be done in the next section.  

With the two parameter sets of a WS potential from Table~\ref{para}, 
we have calculated the $\delta_{RO}$ correction using the formalism
outlined in section~\ref{formalism}. 
We choose only $sd$ shell emitters which, most of them, are well described 
by the so-called universal $sd$ interactions --- USD, USDA/B~\cite{USD,USDab}. 
They include $^{22}\mbox{Mg}$, $^{26m}\mbox{Al}$, $^{26}\mbox{Si}$, $^{30}\mbox{S}$, $^{34}\mbox{Cl}$, $^{34}\mbox{Ar}$,
$^{38m}$K and $^{38}$Ca. Six of these transitions are used to deduce the
most precise $\mathcal{F}t$ value, while the decays of $^{26}\mbox{Si}$ and $^{30}\mbox{S}$ are expected to be
measured with an improved precision in future radioactive-beam facilities. 
The shell-model calculations have been performed in the full $sd$ shell,
using the NuShellX@MSU code~\cite{NuShellX}. 

\begin{figure}[ht!]
\centering
		\includegraphics[scale=1.1]{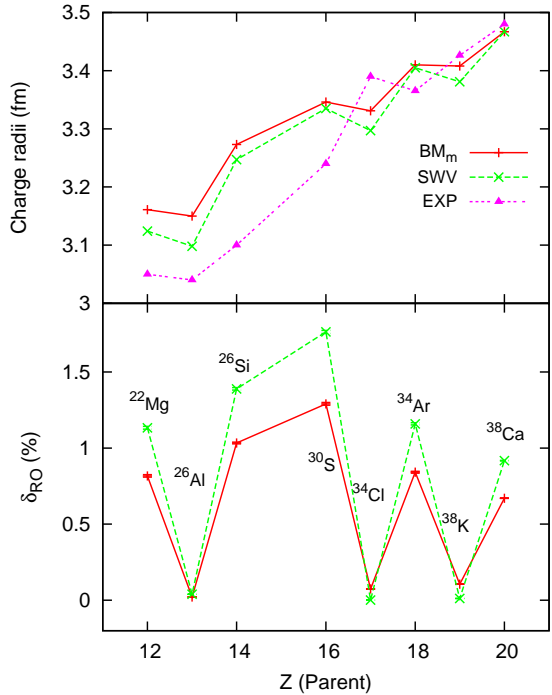}
		\caption{(Color online). Results with the standard parametrizations (BM$_m$ and SWV) of a  WS potential. 
		These results are the averages of three values obtained with the USD, USDA and USDB effective interactions. 
		However, there is a great consensus among these interactions, the resulting uncertainties are negligible. 
		The experimental data on $^{34}$Ar and $^{38}$K are taken from Ref.~\cite{AngMari2013}, while Ref.~\cite{ToHa2002} 
		is used for the others. }
		\label{fig1}
\end{figure}

Fig.~\ref{fig1} shows the results obtained with the two parameter sets. 
The calculated values for charge radii of the parent nuclei as well as 
the values obtained from electron scattering experiments~\cite{AngMari2013} or from isotope-shift estimations~\cite{ToHa2002} 
are plotted in the upper panel. % of Fig.~\ref{fig1}. 
Our calculations of charge radii have been carried out with the formula: 
\begin{equation}\label{ray}
\braket{r^2}_{ch} = \displaystyle \int_0^\infty \rho_p(r) r^4 dr /  \int_0^\infty \rho_p(r) r^2 dr + \frac{3}{2}(a_p^2-b^2/A) \,. 
\end{equation}
The last two terms in this expression correct for the internal structure of the proton and 
for the center-of-mass motion, respectively, 
as detailed in Eq.~\eqref{Rc}. The proton density, $\rho_p(r)$, is defined as 
\begin{equation} 
\rho_p(r) = \frac{1}{4\pi} \sum_\alpha |R_{\alpha_p}(r)|^2 \times n_{\alpha_p}. 
\end{equation}

The occupations $n_{\alpha_p}$ are equal to $(2j+1)$ for fully-filled orbitals of an inert core, 
while for valence orbitals the occupation numbers are obtained from the shell-model diagonalization. 

As seen from Fig.~\ref{fig1}, for nuclei of the lower part of the $sd$ shell, 
including $^{22}$Mg, $^{26m}$Al, $^{26}$Si and $^{30}$S, 
the calculated values of charge radii considerably overestimate the experimental data. 
At the same time, for $^{34}$Cl and $^{38}$Ca our theoretical calculations work rather well. 

Obviously, the $\delta_{RO}$ correction is strongly parametrization dependent, 
as illustrated in the lower panel of Fig.~\ref{fig1}. 
%the numerical values are shown in the last two columns of table~\ref{tab1}. 
With both parameter sets, there appears to be some sort of odd-even staggering. 
Namely, for the parent nuclei with even $Z$ ($N\ne Z$), we obtained a large overlap between proton and neutron radial wave functions, 
thus $\delta_{RO}$ increases. In cases of odd $Z$ ($N=Z$), the overlap is very closed to unity, thus resulting in a very small correction value. 
This effect is solely generated by the isovector terms which also violate the isospin symmetry. The SWV parametrization has a stronger odd-even oscillation, 
indicating that its isospin-dependent term has a stronger effect than that of the other parametrization. 

Note that the parameter sets in Table~\ref{para} were determined by a global fit 
to various ground-state properties of nuclei around doubly-magic nuclei, 
they are generally not adequate for open-shell nuclei far away from the valley of stability, 
including those considered in the present study. 
One could also notice that these parametrizations do not account for the charge-symmetry-breaking effects which have 
been observed in nucleon-nucleon elastic scattering~\cite{Fearing}. 
Furthermore, the charge symmetric isovector term~\eqref{symm} or \eqref{iso} is somehow related 
to the difference between neutron and proton numbers, 
thus can vanish in some particular conditions. Obviously, this latter property 
does not agree with the HF case~\cite{OrBr1985}, because the isovector component of a self-consistent mean field 
can not be identically zero due to the difference between proton and neutron densities.

The WS potential is constructed based on the nuclear phenomenology, 
it does not have a direct connection to the nucleon-nucleon interaction as in the HF case. 
To improve the accuracy of the WS potential, we could add either additional terms or increase the number of parameters. 
For example, one could define differently the form factor for the volume, spin-orbit and isospin-dependent terms. 
However, this idea is always limited, because of the lack of experimental data to constrain new parameters. 
In general, we only have the charge radii and the separation energies which could be predicted by the WS model. 
In what follows, we adopt the simplest strategy, re-adjusted case-by-case the parameter $r_0$ and $V_0$ to reproduce 
the charge radii and the separation energies respectively, while the other parameters are fixed at the standard values. 

According to the Koopman's theorem, 
the energy of the highest occupied orbital is approximately equal to the nucleon separation energy with an opposite sign. 
Therefore, one usually fits the last occupied single-particle state and keeps the same potential
to get all the other radial wave functions. However, this corresponds to the extreme independent-particle model. 
For the present study, we re-adjusted $V_0$ for each valence orbital separately. We believe that this method 
is more consistent with the shell model in which the single-particle states are partly occupied.

%\begin{figure}[ht!]
%\centering
%		\includegraphics[scale=0.9]{figures/r0}
%		\caption{(Color online). 
%}
%		\label{f2}
%\end{figure} 

\begin{figure}[ht!]
\centering
		\includegraphics[scale=1.1]{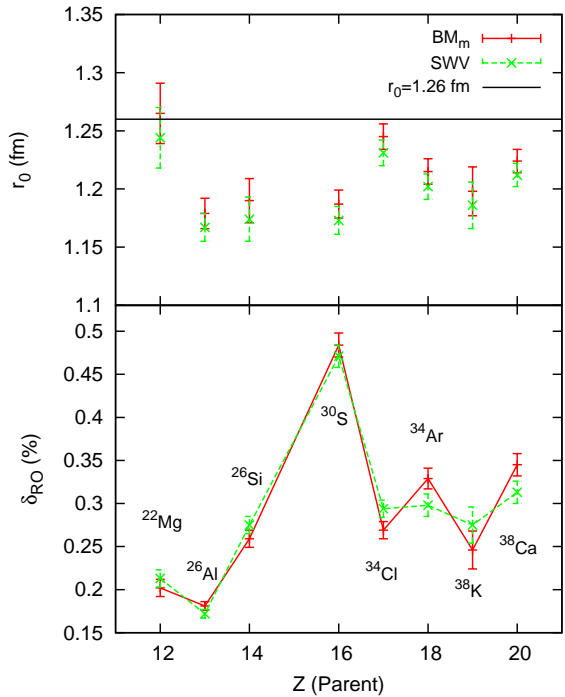}
		\caption{(Color online). Results with re-adjusted parametrizations of WS potentials. 
The length parameter resulting from the fit of charge radii is plotted in the upper panel, 
the horizontal line indicates the standard value, $r_0=1.26$ fm. The correction, $\delta_{RO}$ is plotted in the lower panel. }
		\label{fig2}
\end{figure}

The results obtained from these calculations are illustrated in Fig~\ref{fig2}, from which 
it can be seen that, within the re-adjusted parametrizations, the staggering on $\delta_{RO}$ becomes softer 
and the parametrization dependence is completely removed. The reason is that the fit of separation energies affects the 
original isospin-dependence terms of the selected parametrizations, and brings them to a new term 
determined by the difference between neutron and proton separation energies. 
We recall that in the SWV parametrization, the radii of the potential are expressed with $A$ instead of $(A-1)$, 
that is why the resulting values of $r_0$ are about 1.2\% lower than those 
obtained with the BM$_m$ parametrization. 

Two sources of uncertainties are considered, one is the error on the experimental data of charge radii and 
the other is the spread of results obtained with different shell-model effective interactions. 
We assume that the calculations with different interactions provide a set of independent values, we can 
thus apply statistics to describe this data set. Our adopted values are the normal averages (arithmetic means), 
while the spread of the individual values being considered as a {\it statistical} uncertainty 
that follows a normal (or Gaussian) distribution. 
The uncertainties are dominated by the errors on the experimental charge radii and
they only weakly depend on a particular effective interaction 
and as specific parametrization of the WS potential. 
For this reason, we consider this source of uncertainties as {\it systematic}. 
To cover the small spread, the maximum value has to be chosen. 

For each individual calculation, we compute the charge radii and the radial overlap correction 
for four different values of $r_0$ around $1.26$ fm. 
Both quantities can be very well approximated by linear functions in the vicinity of 1.26~fm.
So, we fit the results by straight lines, e.g.  
\begin{eqnarray}\label{linear}
\sqrt{\braket{r^2}_{ch}} &=& a\times r_0 + b  , \\ \nonumber
\delta_{RO} &=& c\times r_0 + d, 
\end{eqnarray}
where $a$, $b$, $c$ and $d$ are the regression coefficients. %, to be determined by the least-squares fit. 

Once these coefficients are determined, we can deduce the radial overlap correction and the length parameter that correspond 
to the experimental charge radii. To extract the systematic uncertainty on $\delta_{RO}$, 
we followed the error propagation rule,  
\begin{equation}\label{sys}
\sigma_{syst} = \sqrt{ \big( c \times \sigma_{r_0}\big)^2 + \big( r_0\times \sigma_{c} \big)^2
+ \big( \sigma_{d} \big)^2 } , 
\end{equation}
In this equation, $\sigma_{r_0}$ is the systematic uncertainty on the length parameter, 
evaluated from the first line of Eq.~\eqref{linear}, while
$\sigma_c$ and $\sigma_d$ are the errors of the coefficients $c$ and $d$, obtained from the fit. 
Note that, for all cases, the dispersion of the data points around the straight line is almost negligible, 
thus the errors $\sigma_c$ and $\sigma_d$ are generally not significant. %, i.e. $\chi^2/\nu\approx 1$. 

It is important to remark that $\sigma_{syst}$ depends on the sensitivity to the length parameter 
(on $a$ and $c$ values in Eq.~\eqref{linear}) 
which can be varied from nucleus to nucleus. This effect will be discussed further in the next section. 

The overall uncertainty is estimated as the sum in quadrature, 
\begin{equation}
\sigma = \sqrt{\sigma_{syst}^2+\sigma_{stat}^2}, 
\end{equation}
where $\sigma_{stat}$ is the previously mentioned statistical uncertainty. 

\section{Radial Overlap Correction with Full Parentage Expansion}\label{interm} 

\subsection{Formalism}

Although the $\delta_{RO}$ correction is determined by the overlap between the radial wave function 
of the decaying proton in the parent nucleus and that of the resulting neutron in the daughter nucleus, 
both of these particles are bound to the intermediate system with $(A-1)$ nucleons, therefore the structure of 
this nucleus proves to be important as well~\cite{ToHa2002,ToHa2008,OrBr1985}. In the previous section, 
we have taken only the separation energies relative to the ground states. 
Now, we extend our model to include more complete information associated with that system
as was done in Refs.~\cite{ToHa2002,ToHa2008,OrBr1985}.
We expand the $\delta_{RO}$ correction by inserting a complete sum over intermediate states
$\sum_\pi\ket{\pi}\bra{\pi}$ into the transition densities, between creation and annihilation operator. 
Subsequently, Eq.~\eqref{RO} becomes 
\begin{equation}\label{RO1}
\delta_{RO} = \frac{2}{M_F^0} \sum_{\alpha,\pi} \braket{f| a_{\alpha_n}^\dagger |\pi}^T \braket{i| a_{\alpha_p}^\dagger |\pi}^T (1-\Omega_\alpha^\pi), 
\end{equation}
where the matrix elements, $\braket{f| a_{\alpha_n}^\dagger |\pi}^T$ and $\braket{i| a_{\alpha_p}^\dagger |\pi}^T$ are related 
to the spectroscopic amplitudes~\cite{OrBr1985} for neutron and proton pick-up, respectively. 
These quantities can be computed within the shell model using 
an appropriate isospin-invariant effective interaction. 

Next, we assume that the transferred nucleon and the intermediate nucleus are two 
independent objects. Thus, the initial and final states take the following form, 
\begin{equation}
\begin{array}{lll}
\ket{i} = a_{\alpha_p}^\dagger \ket{\pi} & \mbox{and} & 
\ket{f} = a_{\alpha_n}^\dagger \ket{\pi}. 
\end{array}
\end{equation}

Under this assumption, we obtain, 
\begin{equation}\label{separ}
\begin{array}{lll}
E_i = E_\pi + \epsilon_{\alpha_p} & \mbox{and} & 
E_f = E_\pi + \epsilon_{\alpha_n}, 
\end{array}
\end{equation}
where $\epsilon_\alpha$ is the single-particle energy of a valence orbital, 
$E_i/E_f$ is the energy of the initial/final state and $E_\pi$ is the energy of an intermediate state. 
From Eq.~\eqref{separ}, we can extract $\epsilon_\alpha$ from experimental ground-state masses 
and excitation energies of the intermediate nucleus~\cite{AME2012}. Where experimental excitation energies are not
available, we use those predicted by the shell model.

Note that the intermediate state $\ket{\pi}$ and the single-particle states $\ket{\alpha}$ must have quantum numbers 
such that the coupled states are $J^\pi=0^+$ and $T=1$. 

The expression~\eqref{RO1} allows us to go beyond the closure approximation and
take into account the dependence of $\Omega_\alpha^\pi$ on the excitation energies of the
intermediate $(A-1)$ system, $E_\pi$. For each $E_\pi$, we fine-tune our potential so that the individual energies of valence space orbitals match 
experimental proton and neutron separation energies. 

\begin{figure*}[ht!]
\centering
	%\begin{minipage}[b]{1\linewidth}
		\includegraphics[scale=1.4]{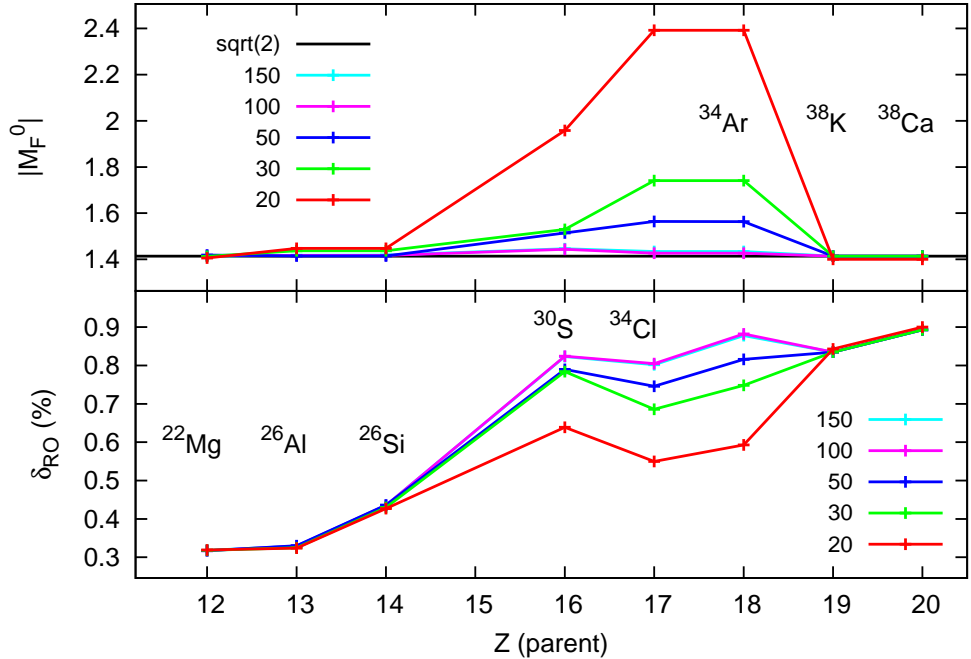}
		\caption{(Color online). Fermi matrix element $|M_F^0|$ and 
		the radial overlap correction $\delta_{RO}$ for various numbers of intermediate states 
		$N_{\pi} \in [20, 150]$.}
		\label{f1}
	%\end{minipage}
\end{figure*}

In general, it is not possible to account for all intermediate states 
because of the computational limit. However, since spectroscopic amplitudes decrease on average with increasing excitation energy $E_\pi$,
one can impose a robust truncation at a certain number of states, $N_\pi$. 
The variations of $|M_F^0|$ and $\delta_{RO}$ 
as a function of $N_{\pi}$ are displayed in Fig.~\ref{f1}. 
From the top panel, it is seen that the Fermi matrix element does not converge quickly. 
For the transitions in the middle of the $sd$ shell, with $N_{\pi}=150$ for each spin and parity,
the value of $|M_F^0|$ is still off its model-independent value. 
Fortunately, the correction $\delta_{RO}$ converges much faster than $|M_F^0|$, because of the factor $(1-\Omega_\alpha^\pi)$ which decreases monotonically 
with increasing of $E_\pi$ and tends to zero finally. 
For all $sd$-shell emitters, one can use $N_\pi$=100 as a reasonable cut-off for the number of intermediate states. 

\subsection{Charge radius calculation}

The parentage-expansion formalism outlined above can be generalized for charge radius calculation. 
The operator for this observable is defined in an occupation number formulation as
\begin{equation}\label{roper}
r_{sm}^2 =\frac{1}{Z} \sum_{\alpha}\bra{\alpha_p} r^2\ket{\alpha_p}a^{\dagger}_{\alpha_p} a_{\alpha_p} \,.
\end{equation}
The subscript $sm$ refers to {\it shell model}, to indicate that the operator, $r_{sm}^2$, 
will be computed using shell-model wave functions. 
The sum in Eq.~\eqref{roper} runs over all single-particle orbitals of the valence space.
We obtain the square of the charge radius (relative to the inert core) as an expectation value of the operator $r_{sm}^2$ 
in the ground state of a parent nucleus\footnote{We diagonalize this operator in the initial $0^+, T=1$ state. 
For most cases, it is the ground state of the parent nuclei, except for $^{26}$Al and $^{38}$K. For these two cases, 
such a state has an excitation energy of 228.3~keV and 130.4~keV, respectively.} : 
\begin{eqnarray}\label{19}
\displaystyle \braket{r^2}_{sm}& = & \bra{\psi_i} r_{sm}^2\ket{\psi_i} \\ \nonumber
\displaystyle &=& \frac{1}{Z} 
\sum_{\alpha}\bra{\alpha_p} r^2 \ket{\alpha_p}\bra{\psi_i}a^{\dagger}_{\alpha_p} 
a_{\alpha_p}\ket{\psi_i} \,.
\end{eqnarray}

Now, we insert the complete sum over intermediate states $\sum_{\pi} \ket{\pi}\bra{\pi}$ into Eq.~\eqref{19}. 
Therefore, the average proton occupancy, $\bra{\psi_i}a^{\dagger}_{\alpha_p} a_{\alpha_p}\ket{\psi_i}$ 
is converted into a product of spectroscopic factors which can be obtained from the shell model. 
Thus, Eq.~\eqref{19} reads,  
\begin{eqnarray}\label{20}
\braket{r^2}_{sm}=\frac{1}{Z} \sum_{\alpha, \pi} \bra{\psi_i}a^{\dagger}_{\alpha_p} \ket{\pi}^2 
\bra{\alpha_p} r^2 \ket{\alpha_p}^\pi .
\end{eqnarray}

The single-particle matrix element is given by
\begin{eqnarray}\label{20}
\bra{\alpha_p} r^2 \ket{\alpha_p}^\pi =  \int_0^\infty r^4 |R_{\alpha_p}^{\pi}(r)|^2  dr  ,
\end{eqnarray}
where the additional label $\pi$ denotes that the radial wave functions 
depend on the intermediate states $\ket{\pi}$ because of the fit of separation energies. 

We have checked the expectation value, $\braket{r^2}_{sm}$, with various values of $N_\pi$. 
We found that this quantity converges much faster than the correction $\delta_{RO}$ or $M_F^0$, 
the value at $N_\pi=50$ is sufficiently accurate. 

%Remark that the operator in Eq.~\eqref{roper} is not optimized for a chosen model space, 
%in principle, an effective operator must be used instead. However, 
%for the sake of simplicity we will use the bare operator. 

Within the shell model exploited here, only a limited number of nucleons in a valence space outside 
an inert core are treated as active nucleons. In this spirit, we calculate the charge radii by two different methods.
Following {\it method I}, we extract the contribution of core orbitals 
from the experimental charge radius, $\braket{r^2}_{ch}^c$ of the closed-shell nucleus (i.e. $^{16}$O for the $sd$ shell) via, 
\begin{eqnarray}\label{core}
\displaystyle\braket{r^2}_{ch} &=&  \braket{r^2}_{sm} + \frac{3}{2} (a_p^2-b^2/A) \\ \nonumber
\displaystyle&& + \braket{r^2}_{ch}^c \bar{Z}  \\ \nonumber
\displaystyle&& + 3/4(2n'+l'+2) (b^2-b_c^2) \bar{Z}  \\ \nonumber
\displaystyle&& - 3/2 (a_p^2-b_c^2/A_c) \bar{Z} , 
\end{eqnarray}
where $\bar{Z}=Z/Z_c$ is the ratio between the atomic number of parent and core nucleus. 
The third line of Eq.~\eqref{core} accounts for the mass-dependence of the potential. We obtained this term 
using harmonic oscillator wave functions (more details of the formalism can be found in Ref.~\cite{Kir2006}). 
The symbols $n'$ and $l'$ stand for the radial and orbital angular momentum quantum numbers 
of the highest filled level of the core, $b_c$ and $A_c$ are the oscillator parameter and 
the mass number, respectively, of the closed-shell nucleus. 
The fourth line of Eq.~\eqref{core} is the center-of-mass correction for the closed-shell nucleus, 
similar to that in Eq.~\eqref{Rc}.  
This method avoids the energy dependence of the nuclear mean field which could be significant for deeply bound states, 
as suggested from the optical model~\cite{LanePRL1962, Lane1962, Hod1971} and 
also from HF calculations using Skyrme forces~\cite{DG}. 

With {\it method II} we calculate the charge radii with WS eigenfunctions for all occupied states, including closed-shell orbits,  
\begin{eqnarray}\label{core1}
\displaystyle\braket{r^2}_{ch} &=&  \braket{r^2}_{sm} + \frac{3}{2} (a_p^2-b^2/A) \\ \nonumber
\displaystyle&& + \frac{1}{Z}\sum_{\alpha} (2j+1) \int_0^\infty r^4 |R_{\alpha_p}(r)|^2 dr, 
\end{eqnarray}
where the sum in the second line of Eq.~\eqref{core1} runs over all inactive orbits below the shell-model valence space. 
Since these orbits are assumed to be fully filled, the proton occupancies are taken as $(2j+1)$. 
We notice that the energy dependence is not accounted for in Eq.~\eqref{core1}, however 
this method is free from the mass-dependent correction which is necessary in the previous method.

We have explored the predictive power of these new approaches for charge radii. 
We found that, with $V_0$ as the only adjustable parameter, the predictive ability of 
our methods I and II is much better than that of the traditional approach, 
except for $^{34}$Cl for which the value obtained from an isotope-shift estimation~\cite{ToHa2002} 
is particularly large.  

As we have mentioned in the introduction, a series of calculations of $\delta_{RO}$ using the shell-model approach 
have been carried out by Towner and Hardy. However, these authors determine the parameter $r_0$ 
by requiring the mean-square radii computed from the traditional approach, Eq.~\eqref{ray}, to match the experimental values, 
while the dependence on intermediate states is not taken into account. 
Then the resulting $r_0$ values are kept for the calculations of $\delta_{RO}$ 
in the full parentage-expansion formalism. 
In the latter step, the depth of the central term is independently re-adjusted to 
reproduce the separation energies 
with respect to multiple-intermediate states. 
We notice that, in principle, the two parameters could not be unambiguously
determined in this way, instead the fit should be performed using the least-squares method  
which ensures the optimization of the resulting radial wave functions. 

\begin{figure}[ht!]
\centering
		\includegraphics[scale=1.1]{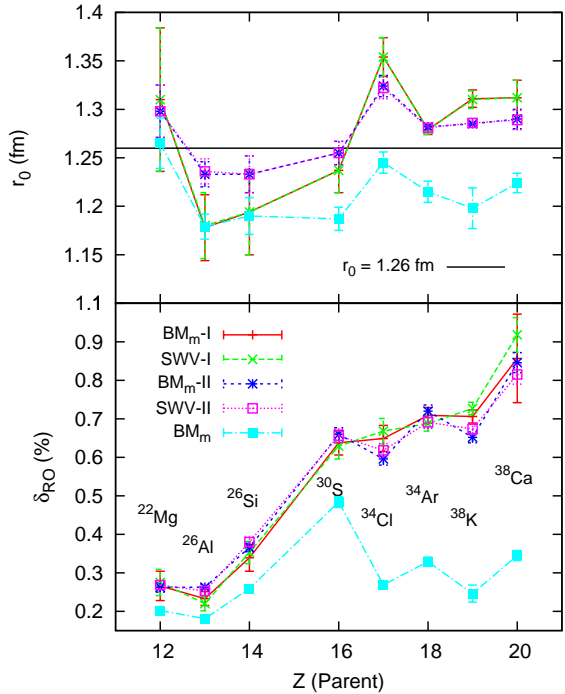}
		\caption{(Color online). Comparison between the results obtained with method I (BM$_m$-I and SWV-I) 
and those obtained with method II (BM$_m$-II and SWV-II). The cyan curve (BM$_m$) represents the values 
obtained with the traditional approach which accounts only for the ground state of the $(A-1)$ nucleus. }
		\label{fig3}
\end{figure}

Thanks to the generalization of the formalism for charge radii described above, %~\eqref{core} and~\eqref{core}, 
we are able now to adjust the potential depth, $V_0$ and the length parameter, $r_0$ 
in a self-consistent way. The final individual energy spectra and wave functions are capable 
thus to reproduce simultaneously the one-proton and one-neutron separation energies and  
the experimental charge radii of the parent nuclei. 
Our results with full parentage expansion are shown in Fig.~\ref{fig3}. 
Although we use two different parametrizations of the WS potential (SWV 
and BM$_m$), when the depth, $V_0$ and the length parameter, $r_0$ are re-adjusted, 
they lead to very similar result. 
Note that the same situation had occurred in our calculations 
with the traditional approach (see discussions in section~\ref{overlap}). 
Moreover, the $\delta_{RO}$ correction does not significantly depend on the treatment 
of closed-shell orbits, it is seen that each method only produces marginally 
different values throughout the $sd$ shell. 
However, the $r_0$ values resulting from method II are closer to 1.26 fm, 
the value obtained from a global fit~\cite{SWV}. 
We found in practice that the fit of charge radii based on method I is generally less appropriate because of low sensitivity. 
Since the closed-shell contribution is taken from the experimental data, 
the charge radii calculated with Eq.~\eqref{core} are almost model-independent, particularly for 
a small number of valence protons. Accordingly, the uncertainties quoted in the BM$_m$-I and SWV-I results  
are more than twice larger than those produced by method II. 
The numerical values of the $r_0$ parameter and of the $\delta_{RO}$ corrections are listed in Table~\ref{tab3}
of Appendix. 

Comparing these results with a result generated by the traditional approach (BM$_m$) 
in which the depth and the length parameter are also re-adjusted but only 
the ground state energy of the $(A-1)$ nucleus is considered, 
it is evident that the introduction of multiple-intermediate states 
has the effect of increasing both the
radial overlap correction and the length parameter, 
particularly for the transitions in the upper part of the $sd$ shell. 

%Since the effect of parametrization dependence is no longer significant, we have used only the BM$_m$ parameter set 
%for our calculation with the method II, denoted as BM$_m$-II. 

%\begin{figure}[ht!]
%\centering
	%\begin{minipage}[b]{1\linewidth}
%		\includegraphics[scale=1.1]{figures/method2}
%		\caption{(Color online). Comparison between the results obtained with the method I (BM$_m$-I and SWV-I) 
%and those with the method II (BM$_m$-II and SWV-II).}
%		\label{fig4}
	%\end{minipage}
%\end{figure}

\subsection{Surface terms}

Although the WS potential is a kind of phenomenological mean field, 
it has its theoretical basis related to the saturation properties of
nuclear matter as discussed in section~\ref{ws}. 
Thus, instead of continuously varying the central part of the potential, it is recommended~\cite{PS} to include an extra surface-peaked term and
adjust its strength to reproduce the nucleon separation energies. Two terms have been considered in the literature~\cite{ToHa2002}, namely,
\begin{equation}\label{g}
V_g(r)=\left(\frac{\hbar}{m_{\pi}c}\right)^2\frac{V_g}{a_sr}\exp(\frac{r-R_s}{a_s}) [ f(r, R_s, a_s) ]^2,
\end{equation}
and
\begin{equation}\label{h}
V_h(r)=V_h\, a_0^2\, \big[ \frac{d}{dr} f(r, R_0, a_0) \big]^2 ,
\end{equation}
where $(\hbar/m_\pi c)\approx 1.4$ fm is the pion Compton wavelength, while $V_g$ and $V_h$ being adjustable parameters. 

\begin{figure}[ht!]
\centering
		\includegraphics[scale=1.15]{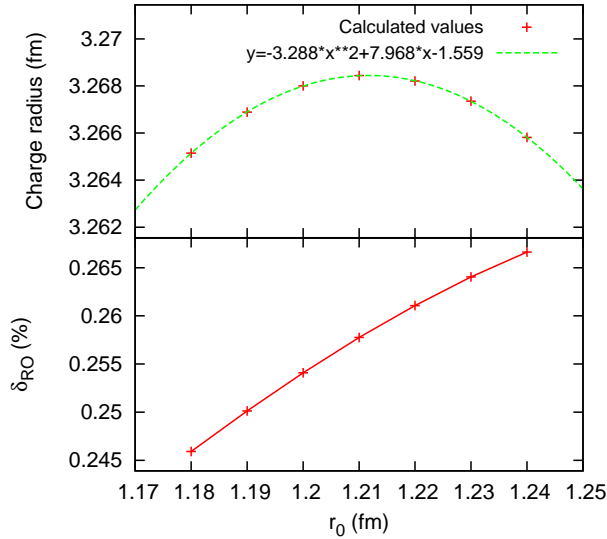}
		\caption{(Color online). Charge radius and $\delta_{RO}$ as a function of the parameter $r_0$ for the case of $^{34}$Cl. 
The calculation has been performed with the BM$_m$ parametrization including the surface term, $V_h(r)$~\eqref{h}. The strength $V_h$ 
was adjusted separately for a proton and a neutron to get the experimental energies relative to various intermediate states, 
whereas the length parameter, $r_0$, was re-adjusted to reproduce the experimental charge radius. }
		\label{fig4}
\end{figure}

However, we found that the term, $V_h(r)$ has a very weak effect on the single-particle spectra, 
the fit of separation energies results thus 
a large value of $V_h$ and generates a high peak on the WS potential at nuclear surface. 
Furthermore, the inclusion of $V_h(r)$ leads to an unusual (quadratic) correlation 
between the charge radius and the length parameter as shown in Fig.~\ref{fig4}. 
This property is in disagreement with the uniform-density liquid drop model~\cite{Gamow}, 
and moreover it deteriorates our optimization procedure. 
For these reasons, we do not use this term for our study of $\delta_{RO}$. 
On the contrary, the term, $V_g(r)$ has a much stronger effect on the single-particle spectra, 
therefore the WS potential complemented with $V_g(r)$ does not 
encounter any particular problem. 

\begin{figure}[ht!]
\centering
		\includegraphics[scale=1.1]{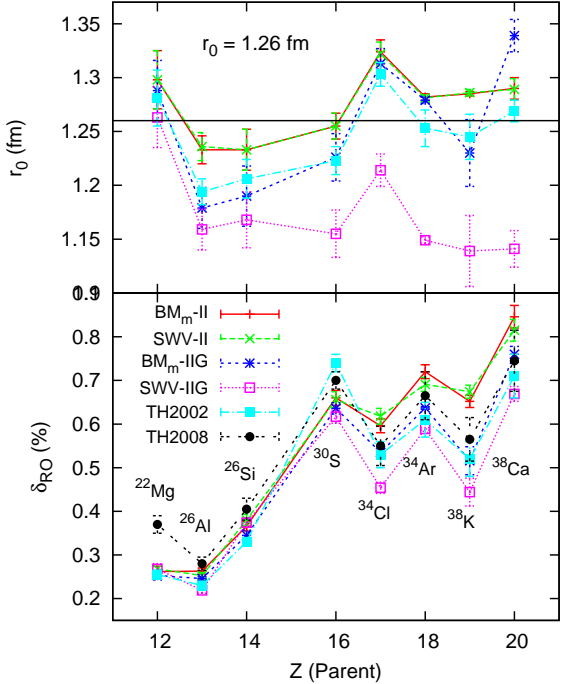}
		\caption{(Color online). BM$_m$-IIG and SWV-IIG are the results for the radial overlap correction 
and the length parameter as obtained based on method II, while the parametrizations (BM$_m$ and SWV) are 
complemented with the surface term $V_g(r)$. Our calculations without $V_g(r)$ term (BM$_m$-II and SWV-II) 
as well as the earlier (TH2002~\cite{ToHa2002}) and the recent (TH2008~\cite{ToHa2008}) 
calculations of Towner and Hardy are shown for comparison. }
		\label{fig5}
\end{figure}

%We have carried out the calculations based on both selected 
%parametrizations of the WS potential with inclusion of $V_g(r)$. 
Before adding the surface term, we fix the depth of the central term, $V_0$, 
in such a way that the calculated energy of the last occupied orbit matches the experimental 
separation energy relative to the ground state of the $(A-1)$ nucleus. 
The energies of the remaining states are fitted 
by varying the strength of the surface term, while the parameter $r_0$ 
is consistently re-adjusted to get the experimental charge radii of the parent nuclei. 
The other parameters are kept fixed at the standard values. 
The results of these calculations are reported in Fig.~\ref{fig5}, with the labels BM$_m$-IIG and SWV-IIG. 

Let us first emphasize the effect of the $V_g(r)$ term on the charge radii. 
With $r_0$ being fixed, they come out to be larger 
and for most cases overestimate the experimental values, especially the calculation with the SWV parametrization  
where the radii of the potential are scaled as $A^{1/3}$ instead of $(A-1)^{1/3}$. 
However, within the BM$_m$ parametrization the calculated charge radii for 
$^{22}$Mg, $^{34}$Cl and $^{38}$Ca are somewhat lower than the experimental values. 
The numerical result of this calculation is reported in Table~\ref{radvg}. 

\begin{table}[h!]
\centering
\caption{Charge radii (in fm units) calculated by method II with inclusion of the surface term, $V_g(r)$, 
while the parameter $r_0$ is fixed at 1.26 fm. The experimental values are listed in the last column.}
\setlength{\extrarowheight}{0.08cm}
\begin{tabular}{ |p{1.3cm} | p{1.7cm}|p{1.7cm}|p{1.15cm}|}
\hline
   Nucleus     &  BM$_m$-IIG  &   SWV-IIG & Exp     \\
\hline
  $^{22}$Mg    & 3.027    &   3.047   & 3.05        \\
  $^{26}$Al    & 3.088    &   3.1     & 3.04 \\
  $^{26}$Si    & 3.143    &  3.159    & 3.1  \\
  $^{30}$S     & 3.26     &   3.309   & 3.24  \\
  $^{34}$Cl    & 3.346    &   3.426   & 3.39  \\
  $^{34}$Ar    &  3.345   &   3.436   & 3.365    \\
  $^{38}$K     & 3.432    &   3.494   & 3.426 \\
  $^{38}$Ca    & 3.417    &  3.56     &  3.48  \\
\hline
\end{tabular}
\label{radvg}
\end{table}

As the radii computed using WS radial wave functions are generally proportional to the length parameter, 
our fit results thus in smaller $r_0$ as seen from Fig.~\ref{fig5}, except the three cases mentioned above. 
Although these calculations produce smaller $r_0$, the $\delta_{RO}$ values 
for the cases with masses between $A=22$ and $30$, are in fair agreement with those obtained 
in the calculations without the $V_g(r)$ term.  
As for the other transitions, the SWV-IIG values of $\delta_{RO}$ are about 23\% lower than the BM$_m$-II or SWV-II values, 
whereas those obtained from the BM$_m$-IIG model drop by about 13\% only. 
Furthermore, in this heavy-mass region, the inclusion of the $V_g(r)$ term leads to a significant dependence on the WS parametrization, 
even though the parameter $V_0$ has been also re-adjusted for the ground state. 
%Obviously, the radial overlap correction is consistent with the length parameter resulted from the fit of charge radii. 
Note that the uncertainties on these latter results are somewhat larger than the uncertainties on the results of method II (see Table~\ref{tab3}), 
that is to say that the sensitivity to $r_0$ becomes lower (the coefficient $c$ in Eq.~\eqref{sys} increases) when we include the $V_g(r)$ term. 

\subsection{Discussion}

Fig.~\ref{fig5} shows the comparison of the present results with
the recent (TH2008, Ref.~\cite{ToHa2008}) and earlier (TH2002, Ref.~\cite{ToHa2002}) results of 
Towner and Hardy. 
The latter correspond to the values tabulated in the last column 
of Table V in Ref.~\cite{ToHa2002}, where the shell-model calculations 
for nuclei between $A=22$ and $34$ have been done in the full $sd$ 
shell and for those with $A=38$, the model space included the $0f_{7/2}$ 
orbital while the $0d_{5/2}$ orbital being frozen. 
In 2008~\cite{ToHa2008}, these authors reported that the 
core polarization has a significant effect on $\delta_{RO}$. 
%in particular the transitions in the vicinity of the cross shells. 
Since then they evaluate the radial overlap correction with the inclusion of 
the orbitals outside the valence space, their method is based on shell-model calculations
of the spectroscopic amplitudes, but limits the sums over single-particle 
orbitals to those for which large spectroscopic factors have been observed in neutron pick-up reactions. 
We do not consider this effect in the present study, our work in this direction is in progress. 

With the exception of $^{30}$S, the BM$_m$-II or SWV-II values for $\delta_{RO}$ are on average 12\% larger 
than those of TH2002. 
This augmentation could be understood as due to the increase of $r_0$ because our calculation
takes into account all intermediate states for the charge radius. 
Perhaps, for the cases of $A=38$, 
this effect is partly due to the inclusion of the $0f_{7/2}$ orbital in the TH2002 calculation. 
The fact that Towner and Hardy obtained a larger value for $^{30}$S may 
rely on the cut-off for the sum over intermediate states (this information is not stated
in their articles). 
Although a full $sd$-shell model space calculation is, at present, 
feasible for all nuclei with $8 \le N,Z \le 20$, the convergence of one hundred 
states in the mid-shell region is still time-consuming.
The values obtained from the recent calculations of Towner and Hardy (TH2008) 
are somewhat larger than their earlier result (TH2002), especially for $^{22}$Mg. 
This may be due to the different model spaces used as we have mentioned above. 
It can be seen from Fig.~\ref{fig5} that the $\delta_{RO}$ values resulting 
from our SWV-IIG calculation follow a very similar trend 
to that of the TH2008 values, but they are about 16\% lower in magnitude. 
One could quickly guess that these two sets of $\delta_{RO}$ will produce 
a very similar agreement with the CVC hypothesis, but with different $\mathcal{F}t$ values. 

Note that the results of Towner and Hardy are obtained from their assessment 
of all multiple-parentage calculations made for each decay, 
including the calculation without additional terms and 
the calculation with the $V_h(r)$ and $V_g(r)$ terms.  
However, each of these calculations produced very similar 
values of $\delta_{RO}$ because they used the 
same set of the length parameter which are determined using 
a traditional method (see discussions in Ref.~\cite{ToHa2002}). 

We notice that the calculation with a surface-peak term could be very dependent on the fitting procedure. 
For example if one fixes the depth of the volume term ($V_0$) the same for the proton and the neutron, then 
one adjusts the parameter $V_g$ and $r_0$ to reproduce the relevant experimental observables, 
the conventional isovector terms presented in the central part of the potential 
will not or only weakly be affected by this optimization procedure because of the difference between the form factor 
of the volume and the surface term. 
Consequently, the resulting $\delta_{RO}$ will show a stronger odd-even staggering as we have seen in section~\ref{overlap}. 

\section{Constancy of the $\mathcal{F}t$ values}\label{Const}

With our results, we are now in a position to check the constancy of the $\mathcal{F}t$ values, the criteria to validate the CVC hypothesis. 
The individual $\mathcal{F}t$ values computed using the expression~\eqref{1}  
are averaged (denoted as $\overline{\mathcal{F}t}$) and listed in column 2 of Table~\ref{t2}. 
The input data for $ft$, $\delta_{NS}$, $\delta_R'$ and $\delta_{IM}$ 
are taken from Ref.~\cite{HaTo2015}. For each average, we also compute the $\chi^2/\nu$,  
which measures the scatter of the individual $\mathcal{F}t$ values relative to the mean. 
Here $\nu$ is the number of degrees of freedom, equal to $N-1=5$. 
The decays of $^{26}$Si and $^{30}$S are not included because 
of large experimental uncertainties on their experimental $ft$ values. 
Then the scale factor $s=\sqrt{\chi^2/\nu}$ is used to establish the quoted uncertainty on $\overline{\mathcal{F}t}$. 
The statistics procedure followed here is that recommended by 
the Particle Data Group~\cite{PDG}. 

From their latest survey~\cite{HaTo2015}, Hardy and Towner 
did not include any uncertainties on the $\delta_R'$ correction, 
but treated the contribution of the $Z^2\alpha^3$ term in $\delta_R'$ 
as a source of systematic uncertainty, to be assigned to $\overline{\mathcal{F}t}$. 
In the present calculations, we adopt from that survey, adding into $\overline{\mathcal{F}t}$ a systematic uncertainty of 
$\pm$0.36 sec.\footnote{To simplify, we take this value directly from Ref.~\cite{HaTo2015}. Regarding their 
procedure, such uncertainty could depend on the sample size and on the calculated $\delta_{RO}$ values.}  
which correspond to the contribution of the $Z^2\alpha^3$ term. 

\begin{table}
\caption{Reported in the left half (column 2 to 4) are the averages, $\overline{\mathcal{F}t}$, 
and the corresponding $\chi^2/\nu$ and CL values, 
while the right half (column 5 to 7) contains the values obtained from similar procedure but without regarding 
the theoretical uncertainties on $\delta_{RO}$. 
In order to reduce the table width, we labeled the model for each theoretical calculation 
from A to H: A=BM$_m$-I, B=SWV-I, C=BM$_m$-II, D=SWV-II, E=BM$_m$-IIG, F=SWV-IIG, 
G=TH2002, H=TH2008. 
  }
\setlength{\extrarowheight}{0.08cm}
\begin{tabular}{ l | lll | lll | }
%\hline 
%\multicolumn{13}{|c|}{Results for the 6 best-known emitters under consideration ($\nu=5$)} \\
\cline{2-7} 
& \multicolumn{3}{c|}{With uncer. of $\delta_{RO}$} & \multicolumn{3}{c|}{No uncer. of $\delta_{RO}$} \\
\hline 
\multicolumn{1}{|c|}{Model} & $\overline{\mathcal{F}t}$ & $\chi^2/\nu$ & CL & $\overline{\mathcal{F}t}$ & $\chi^2/\nu$ & CL \\
\hline 
  \multicolumn{1}{|c|}{A}    & 3070.1(15)  & 5.09  & 0    &  3070.6(15) & 6.97 & 0      \\
   \multicolumn{1}{|c|}{B}   & 3070.4(18)  & 6.45  & 0    &  3070.4(17) & 8.94 & 0      \\
  \multicolumn{1}{|c|}{C}  & 3071.2(10) & 2.84  & 1    &  3071.1(10)& 3.08 & 0      \\
   \multicolumn{1}{|c|}{D}	    & 3071.0(13)  & 3.93  & 0    &  3070.8(13) & 4.35 & 0      \\
   \multicolumn{1}{|c|}{E} & 3072.90(70) & 1.06  & 38   &  3072.76(70)& 1.25 & 28     \\
   \multicolumn{1}{|c|}{F}    & 3074.49(80) & 0.46  & 81   &  3074.45(70)& 0.49 & 78    \\
   \multicolumn{1}{|c|}{G}	    & 3072.84(80) & 1.92  & 9    &  3072.75(80)& 2.02 & 7      \\
   \multicolumn{1}{|c|}{H}	    & 3072.26(90) & 0.57  & 72   &  3071.93(70)& 0.82 & 54     \\
\hline 
\end{tabular}
\label{t2}
\end{table} 

From the obtained values of $\chi^2$, we could proceed to calculate the confidence level (CL) or the $p$ value, which is defined as 
\begin{equation}
p = \int_{\chi^2_0}^\infty P_\nu (\chi^2) d\chi^2, 
\end{equation}
where $P_\nu (\chi^2)$ is the $\chi^2$ distribution function and $\chi^2_0$ denotes the values computed 
with the null hypothesis (in our case, CVC is the null hypothesis). The calculated CL values  
for each model are given in Table~\ref{t2}. 

%Let us first make a remark on results for the weighted averages. 
Method I produces the smallest values of $\overline{\mathcal{F}t}$, with the highest $\chi^2/\nu$. 
Under the assumption that CVC is valid, these results are statistically significant at CL < 1\%. 
We believe that this discrepancy reflects the inaccuracy 
of the $\delta_{RO}$ values generated by this method 
because of the sensitivity problem, as discussed in the previous section. 
Along these lines, the BM$_m$-I and SWV-I calculations must definitely be rejected. 
Concerning the results of method II, the agreement with CVC is 
somewhat better, but still significantly poorer than the two results of Towner and Hardy. 
In contrast, the calculations which include the $V_g(r)$ term represent the best model for generating 
a set of $\delta_{RO}$ corrections, satisfying the CVC hypothesis. 
The values resulting from the  BM$_m$-IIG calculation are of similar quality to those of TH2002, 
whereas the SWV-IIG calculation produce an even better result and are comparable to that of TH2008. 

In order to assess the constancy of the $\mathcal{F}t$ values from 
the 8 sets of $\delta_{RO}$ on an equal footing, 
we perform a parallel analysis, by setting for all models, 
the theoretical uncertainties on $\delta_{RO}$ to be equal to zero. 
The outcome is given in the right part of Table~\ref{t2}, column 5 to 7. 
It is seen that the omission of this source of uncertainties only slightly affects the weighted averages.   
The $\chi^2/\nu$ values are systematically increased, 
resulting thus in a lower confidence level. Nevertheless, the conclusions of a comparative analysis
of various methods remain unchanged. 

However, it might be too early to draw any conclusion about the Standard Model because 
our samples are made up of only 6 out of the 14 best-known superallowed transitions. 
Our purpose is rather to provide, 
at least qualitatively, an alternative assessment  
for our theoretical models, and to compare with the previous calculations.

\section{CVC test for $\delta_{RO}$ correction}\label{CVC}

In this section, we carry out the confidence-level test proposed recently by 
Towner and Hardy~\cite{ToHa2010}, taking into account the experimental uncertainties, 
as well as uncertainties on $\delta_{RO}$ 
and the other theoretical correction terms. %~\cite{HaTo2015}.
The test is based on the assumption that the CVC hypothesis
is valid to at least $\pm0.03\%$, which is the level of precision currently attained by
the best $ft$-value measurements. 
This implies that a set of structure-dependent corrections should produce a 
statistically consistent set of $\mathcal{F}t$ values. 

If we assume that the CVC hypothesis is satisfied ($\mathcal{F}t$ is constant), 
without regarding the CKM unitarity, we can convert those experimental $ft$ values into experimental 
values for structure-dependent corrections and compare the results with each theoretical calculation in turn. 
Since the isospin mixing correction $\delta_{IM}$ is small compared to the radial overlap correction $\delta_{RO}$ 
and only one set of calculated $\delta_{NS}$ correction exists \cite{ToHaRL1994}, 
pseudo-experimental values for $\delta_{RO}$ can thus be defined by 
\begin{equation}\label{exp}
\delta_{RO}^{ex} = 1+\delta_{NS}-\delta_{IM} -\frac{{\mathcal{F}t}}{ft(1+\delta_R')} \, .
\end{equation}

To test a set of radial overlap correction for $N$ superallowed transitions, 
we use the method of least squares with $\mathcal{F}t$ as the adjustable parameter, 
to optimize the agreement with the pseudo-experimental values: 
\begin{equation}\label{chi}
\displaystyle\chi^2/\nu = \displaystyle \frac{1}{N-1} 
\sum_i^N \frac{ [\delta_{RO}^{th}(i) - \delta_{RO}^{ex}(i) ]^2 }{ \sigma_{th}(i)^2 + \sigma_{ex}(i)^2 }, 
\end{equation}
where $\sigma_{ex}$ and $\sigma_{th}$ stand for the uncertainties on 
the experimental and calculated values of $\delta_{RO}$ respectively. 
The former is propagated from the right-hand side of Eq.~\eqref{exp}, 
based on the data of $ft$, $\delta_{NS}$ and $\delta_R'$ taken from Ref.~\cite{HaTo2015}. 
%From their latest survey~\cite{HaTo2015}, Hardy and Towner suggested that the contribution of the $Z^2\alpha^3$ term 
%in $\delta_R'$ should be treated as a source of systematic uncertainty, assigning to the weighted average, $\overline{\mathcal{F}t}$. 
%While they did not include any uncertainties on the individual $\delta_R'$. 
%Therefore, we propagate $\sigma_{ex}$ with zero uncertainty from $\delta_R'$. 

\begin{table}
\caption{Similar to the results given in Table~\ref{t2}, except that $\mathcal{F}t$ 
is treated as an adjustable parameter. We added the subscribe $min$ into $\chi^2/\nu$ 
to indicate the minimal or the optimized values. 
The corresponding $\mathcal{F}t$ values are referred to as renormalized values 
and denoted as $\mathcal{F}t_R$. 
The values listed in the left part are resulted from the analysis that includes theoretical 
uncertainties on $\delta_{RO}$, whereas those given in the right part are obtained 
without regarding this uncertainty source. }
\setlength{\extrarowheight}{0.08cm}
\begin{tabular}{ l | lll | lll |}
\cline{2-7} 
& \multicolumn{3}{c|}{With uncer. of $\delta_{RO}$} & \multicolumn{3}{c|}{No uncer. of $\delta_{RO}$} \\
\hline 
  \multicolumn{1}{|c|}{Model} & $\mathcal{F}t_R$ & $[\chi^2/\nu]_{min}$ & CL & $\mathcal{F}t_R$ & $[\chi^2/\nu]_{min}$ & CL \\
\hline 
  \multicolumn{1}{|c|}{A}	& 3067.43 & 0.14 & 98      & 3067.48 & 0.17 & 97 \\
   \multicolumn{1}{|c|}{B}	& 3066.81 & 0.10 & 99      & 3066.86 & 0.11 & 99 \\
  \multicolumn{1}{|c|}{C}	& 3069.09 & 0.17 & 97      & 3069.09 & 0.17 & 97 \\
   \multicolumn{1}{|c|}{D}	& 3068.45 & 0.19 & 97      & 3068.45 & 0.20 & 96 \\
   \multicolumn{1}{|c|}{E}      & 3071.82 & 0.25 & 94      & 3071.71 & 0.28 & 92 \\
   \multicolumn{1}{|c|}{F}	& 3074.18 & 0.26 & 93      & 3074.12 & 0.27 & 93 \\
   \multicolumn{1}{|c|}{G}	& 3071.22 & 0.41 & 84      & 3071.35 & 0.47 & 80 \\
   \multicolumn{1}{|c|}{H}	& 3071.13 & 0.22 & 95      & 3071.11 & 0.27 & 93 \\
\hline 
\end{tabular}
\label{t3}
\end{table} 

Thus, the success of each theoretical calculation can be judged by the quality of the fit. 
The result for the renormalized $\mathcal{F}t$: $\mathcal{F}t_R$, 
the optimized $\chi^2/\nu$: $[\chi^2/\nu]_{min}$ and 
the corresponding CL values are given in columns 2 to 4 of Table~\ref{t3}, 
while the values obtained without uncertainties on $\delta_{RO}$ 
are reported in columns 5 to 7. 
From both results, all 8 sets of $\delta_{RO}$ 
(including those generated by the method I) come out to be 
greatly consistent with the CVC hypothesis with the optimized values of $\chi^2/\nu$ ranging from $0.1$ to $0.4$ 
and the confidence level being greater than $80$~\%. 
However, there is a significant spread among model calculations in the deduced $\mathcal{F}t_R$ values. 
It is seen that with the exceptions of the BM$_m$-IIG and SWV-IIG models, the $\mathcal{F}t_R$ values are 
about 3~s lower than the weighted averages, $\overline{\mathcal{F}t}$, given in Table~\ref{t2}. 

From these results, we conclude that the statistical analysis of this section has very low comparative power.  
The result given in Table~\ref{t3} is not accurate enough to make a clear 
decision on selecting one of the theoretical models. 
This indicates that the $\chi^2$ test, Eq.~\eqref{chi}, is not sensitive to small spreads between the correction sets.
Obviously, although method I has been found to be inappropriate, 
the present analysis yields a good agreement of these correction values with the CVC hypothesis,
comparable to the other calculations summarized in Table~\ref{tab3}.
It is likely that a weak sensitivity of the $\chi^2$ test based on Eq.~\eqref{chi} is due to the 
small number of transitions considered here and the result should be reconsidered when more
emitters are included.

To close this section, two remarks can be done: 
\begin{itemize}
\item The shell-model configuration space for the cases with $A=38$ is relatively small. 
These nuclei, having two holes coupled to the inert $^{40}$Ca core, 
have been the subject of recent interest, both experimental and theoretical. 
Results of several theoretical calculations have emphasized the inadequacy of 
the shell model to explain their structure if the configuration is limited to the $sd$ shell. 
One of the most explicit example is that such a model space cannot generate negative-parity states. 
We expect that the agreement of our results with CVC will be improved 
if the configuration space is extended to cover the lowest $pf-$shell orbitals. 
\item It is astonishing that our calculations that include the surface term, $V_g(r)$, 
are in very good agreement with the CVC hypothesis. 
It is demonstrated in the previous section that, with fixed $r_0$, these calculations failed in 
reproducing the experimental charge radii, especially the SWV-IIG model. 
%Without worrying about this 
%problem, we might conclude that the BM$_m$-II and SWV-II models have overestimated 
%the radial overlap correction for the transitions in the upper$-sd$ shell. 
\end{itemize}

%We realize that the $\chi^2$ is sensitive to sample size. 
%The distribution is highly skewed for small values of $\nu$, and becomes more symmetric as $\nu$
%increases, approaching a Gaussian distribution for large $\nu$. 
%In general, a larger sample sizes will give more reliable results with greater precision. 

\section{Summary and perspectives}\label{Summary}

We have performed a detailed and critical study of the radial overlap correction, 
which is the major part of the isospin-symmetry-breaking correction to superallowed $0^+ \to 0^+$ $\beta$ decay.
8 emitters in the $sd$ shell have been re-examined, 
using the USD, USDA and USDB effective interactions, while the single-particle matrix elements of the transition operator 
are calculated with WS eigenfunctions. 

We have investigated two WS potential parametrizations with different isovector terms,
optimizing them in a two-parameter grid $(r_0,V_0)$ to experimental nuclear charge radii and
nucleon separation energies. As a new feature, we have introduced a parentage expansion
to the nuclear charge radius, allowing us to perform a consistent adjustment of both parameters. 
All results have been thoroughly studied with respect to convergence 
as a function of the number of intermediate states. 
Two different approaches to nuclear charge radii with respect to the treatment of closed-shell orbitals 
and two different choices for adjusting the WS potential (variation of the central or surface term)
led us to propose a set of six calculations of the correction for $sd-$shell nuclei. 
However, two calculations have been found to be inappropriate because of their low sensitivity 
when treating the contribution of closed-shell orbits as a constant, taken 
from experimental radii of closed-shell nuclei. 
We found that the surface term, $V_h(r)$, is not compatible with our consistent adjustment, the reason 
is that this term has a very small effect on single-particle spectra. 

For $^{22}$Mg, $^{26}$Al and $^{26}$Si, 
the results on $\delta_{RO}$ obtained stayed close to those obtained by Towner and Hardy in 2002,
where the same model space was exploited. All of our models produced smaller values for $^{30}$S, 
we suppose that this discrepancy is due to the difference in the cut-off for intermediate states. 
In the cases of $^{34}$Cl, $^{34}$Ar, $^{38}$K and $^{38}$Ca, the correction is strongly 
dependent on the method for fitting the experimental data. 

The calculated correction, $\delta_{RO}$, combined with the radiative
corrections ($\delta_R'$ and $\delta_{NS}$) and experimental $ft$ values as surveyed in Ref.~\cite{HaTo2015} 
leads to six new sets of corrected $\mathcal{F}t$ values. 
Most of these values are not concordant with the weighted averages for the six data points 
with a confidence level of $\sim0$\%. Nevertheless, the scatter is much reduced 
for the values resulting from the calculations that include the surface term, $V_g(r)$, the calculation 
based on the BM$_m$ parametrization has a confidence level of 38\% and within the SWV parametrization we obtained a confidence level of 81\%. 

Within the assumption that CVC is valid, we performed the analysis considering the $\mathcal{F}t$ value
as an adjustable parameter and
minimizing the scatter between the calculated values of $\delta_{RO}$ and the pseudo-experimental values. 
This analysis shows that all sets of the correction generated by WS eigenfunctions agree well with the CVC hypothesis. 
However, it is most likely due to the lack of emitters under consideration.

It will be interesting to perform a similar study of lighter and heavier $0^+ \to 0^+$ emitters, 
as well as to enlarge the model space
for nuclei near the cross shell using large-scale calculations.
The aim is to explore the sensitivity of the results to details of the theoretical method and 
to robustly assign the corresponding uncertainties.
The importance stems from the relevance for the most accurate tests of the 
Standard Model of electroweak interactions. 

%An important direction for future research is to construct an effective operator for both Fermi transition and charge radius. 
%The bare operators used in the present study may not be appropriate for a restricted shell-model space. 

\begin{acknowledgments}
We express our appreciation to B. Blank for his stimulating interest to this work,
as well as to M. Bender for several interesting comments and suggestions to theoretical part
and to T. Kurtukian-Nieto for her help with the statistical analysis. 
We also thank B. Blank and T. Kurtukian-Nieto for their careful reading of the manuscript and
a number of useful comments.
L. Xayavong would like to thank the University of Bordeaux for a Ph.D. fellowship. 
The work was supported by IN2P3/CNRS, France.
\end{acknowledgments}

\appendix
\section{Numerical results}

\begin{table*}[ht!]
\caption{Results of the calculations with full parentage expansion 
are tabulated with BM$_m$-I, SWV-I, BM$_m$-II, SWV-II, BM$_m$-IIG 
and SWV-IIG (see section~\ref{interm} for detail). 
Results obtained in our preliminary study, which did not include 
the multiple-intermediate states are denoted by BM$_m$ and SWV. 
These results correspond to those illustrated in Fig.~\ref{fig2}. 
The values taken from Ref.~\cite{ToHa2002} and from Ref.~\cite{ToHa2008} 
(with partial updates from Ref.~\cite{HaTo2015}) are reported with the label TH2002 and TH2008 respectively. }
\setlength{\extrarowheight}{0.08cm}
\begin{tabular}{p{0.5cm}|p{1.4cm}|p{1.4cm}|p{1.4cm}|p{1.4cm}|p{1.4cm}|p{1.4cm}|p{1.4cm}|p{1.4cm}|p{1.4cm}|p{1.4cm}| }
\cline{2-11} 
   & \multicolumn{2}{c|}{BM$_m$-I}  &  \multicolumn{2}{c|}{SWV-I}  & \multicolumn{2}{c|}{BM$_m$-II}  & \multicolumn{2}{c|}{SWV-II} & \multicolumn{2}{c|}{BM$_m$-IIG}  \\
\hline 

\multicolumn{1}{|c|}{$Z$}  & $r_0$ (fm)  & $\delta_{RO}$ (\%) & $r_0$ (fm)  & $\delta_{RO}$ (\%)  & $r_0$ (fm) & $\delta_{RO}$ (\%) & $r_0$ (fm) & $\delta_{RO}$ (\%) & $r_0$ (fm) & $\delta_{RO}$ (\%)  \\
\hline
\multicolumn{1}{|c|}{$12$}	&	1.310(74)&	0.266(38)	&1.310(74)	&0.275(34)	&	1.298(27)&	0.262(12)&	1.298(27)&	0.268(10)& 1.288(48) &0.253(17)	\\
\multicolumn{1}{|c|}{$13$}	&	1.178(34)&	0.233(18)	&1.180(34)	&0.220(19)	&	1.233(13)&	0.263(7)	&	1.236(13)&	0.253(7)	 & 1.179(33) &0.245(9)	\\
\multicolumn{1}{|c|}{$14$}	&	1.194(44)&	0.339(35)	&1.194(44)	&0.353(29)	&	1.233(19)&	0.366(11)&	1.233(19)&	0.380(13)& 1.190(48) &0.345(21) \\
\multicolumn{1}{|c|}{$16$}	&	1.237(23)&	0.638(32)	&1.237(23)	&0.629(33)	&	1.255(12)&	0.660(17)&	1.255(12)&	0.656(18)& 1.226(31) &0.637(21) \\
\multicolumn{1}{|c|}{$17$}	&	1.354(20)&	0.649(34)	&1.354(20)	&0.668(33)	&	1.324(11)&	0.596(16)&	1.322(11)&	0.618(18)& 1.314(25) &0.536(25) \\
\multicolumn{1}{|c|}{$18$}	&	1.278(5)&	0.708(20)	&1.278(5)	&0.686(20)	&	1.282(3)&	0.720(16)&	1.281(3)	&	0.691(16)& 1.280(11)  &0.636(11) \\
\multicolumn{1}{|c|}{$19$}	&	1.302(5)&	0.680(11)	&1.306(8)	&0.714(21)	&	1.285(2)&	0.652(14)&	1.286(3)&	0.674(15)& 1.252(7) &0.538(8) \\
\multicolumn{1}{|c|}{$20$}	&	1.304(16)&    0.889(42)		&1.304(16)	&0.869(46)	&	1.290(10)&	0.846(26)&	1.289(10)&	0.815(25)& 1.341(29) &0.761(39) \\
\hline
& \multicolumn{2}{c|}{SWV-IIG}  &  \multicolumn{2}{c|}{TH2002}  & \multicolumn{2}{c|}{TH2008}  &  \multicolumn{2}{c|}{BM$_m$} &   \multicolumn{2}{c|}{SWV} \\
\hline
\multicolumn{1}{|c|}{$Z$}  & $r_0$ (fm)  & $\delta_{RO}$ (\%) & $r_0$ (fm)  & $\delta_{RO}$ (\%)  & $r_0$ (fm) & $\delta_{RO}$ (\%) & $r_0$ (fm) & $\delta_{RO}$ (\%) & $r_0$ (fm) & $\delta_{RO}$ (\%)  \\
\hline
\multicolumn{1}{|c|}{$12$}&1.263(48)&0.268(21)  &1.281(26) & 0.255(10) &  & 0.370(20) & 1.265(26) & 0.202(10) & 1.244(26)& 0.213(10)  \\
\multicolumn{1}{|c|}{$13$}&1.159(33)&0.219(13)  &1.194(12) & 0.230(10) &  & 0.280(15) & 1.179(13) & 0.181(5)  & 1.167(12)& 0.172(5) \\
\multicolumn{1}{|c|}{$14$}&1.168(46)&0.374(18)  &1.206(18) & 0.330(10) &  & 0.405(25) & 1.190(19) & 0.259(10) & 1.174(19)& 0.275(10) \\
\multicolumn{1}{|c|}{$16$}&1.155(31)&0.616(25)  &1.223(13) & 0.740(20) &  & 0.700(20) & 1.187(12) & 0.484(14) & 1.173(12)& 0.471(13)  \\
\multicolumn{1}{|c|}{$17$}&1.214(26)&0.454(22)  &1.303(11) & 0.530(30) &  & 0.550(45) & 1.245(11) & 0.269(10) & 1.231(11)& 0.294(10) \\
\multicolumn{1}{|c|}{$18$}&1.149(6) &0.587(18)  &1.253(17) & 0.610(40) &  & 0.665(55) & 1.215(11) & 0.329(12) & 1.202(11)& 0.298(13)\\
\multicolumn{1}{|c|}{$19$}&1.162(9) &0.465(12)  &1.245(21) & 0.520(40) &  & 0.565(50) & 1.198(21) & 0.246(22) & 1.186(20)& 0.275(21)\\
\multicolumn{1}{|c|}{$20$}&1.140(31)&0.670(36)  &1.269(10) & 0.710(50) &  & 0.745(70) & 1.224(10) & 0.345(13) & 1.212(10)& 0.313(13) \\
\hline
\end{tabular}
\label{tab3}
\end{table*}

\end{document}